# Venus: The Making of an Uninhabitable World

*Stephen R. Kane (UC Riverside), Giada Arney (NASA GSFC), David Crisp (JPL), Shawn Domagal-Goldman (NASA GSFC), Lori S. Glaze (NASA GSFC), Colin Goldblatt (University of Victoria), Adrian Lenardic (Rice University), Cayman Unterborn (Arizona State University), Michael J. Way (NASA GISS)*

**Abstract**

The goals of the astrobiology community are focussed on developing a framework for the detection of biosignatures, or evidence thereof, on objects inside and outside of our solar system. A fundamental aspect of understanding the limits of habitable environments and detectable signatures is the study of where the boundaries of such environments can occur. Thus, the need to study the creation, evolution, and frequency of hostile environments for habitability is an integral part of the astrobiology story. These provide the opportunity to understand the bifurcation, between habitable and uninhabitable. The archetype of such a planet is the Earth's sister planet, Venus, and provides a unique opportunity to explore the processes that created a completely uninhabitable environment and thus define the conditions that can rule out bio-related signatures. We advocate a continued comprehensive study of our sister planet, including models of early atmospheres, compositional abundances, and Venus-analog frequency analysis from current and future exoplanet data. Moreover, new missions to Venus to provide in-situ data are necessary.

## 1. Studying the Venusian Environment is Imperative for Astrobiology

The prime focus of astrobiology research is the search for life elsewhere in the universe, and this proceeds with the pragmatic methodology of looking for water and Earth-like conditions. In our solar system, Venus is the most Earth-like planet, yet at some point in planetary history there was a bifurcation between the two: Earth has been continually habitable since the end-Hadean, whereas Venus became uninhabitable. Indeed, Venus is the type-planet for a world that has transitioned from habitable and Earth-like conditions, through the inner edge of the Habitable Zone (HZ); thus it provides a natural laboratory to study the evolution of habitability. If we seek to understand habitability, proper understanding of the boundaries of the HZ are necessary: further study and development of our understanding of the evolution of Venus' environment is imperative. Furthermore, current and near-future exoplanet detection missions are biased towards close-in planets (see Section 4), so the most suitable targets for the *James Webb Space Telescope (JWST)* are more likely to be Venus-like planets than Earth-like planets. Incomplete understanding of the evolution of Venus' atmosphere and its present state will hinder the interpretation of these observations, motivating urgent further study.

## 2. The Current Venus: An Uninhabitable Hellscape

Venus could be considered an "Earth-like" planet, because it has a similar size and bulk composition. However, it has a 92 bar atmosphere consisting 96.5% $CO_2$ and 3.5% $N_2$, and a surface temperature of 735 K. Venus' atmosphere is explained by a runaway greenhouse having occurred in the past (Walker 1975), when insolation exceeded the limit on outgoing thermal radiation from a moist atmosphere (Komabayashi 1967; Ingersoll 1969; Nakajima et al. 1992; Goldblatt & Watson 2012; Goldblatt et al. 2013), evaporating the ocean. It is unclear whether the ocean condensed, then later evaporated, or never condensed after accretion (Hamano et al. 2013, H2013). In either case, water loss by hydrogen escape followed, evident in high D/H relative to

Earth (Donahue 1982). Complete water loss would take a few hundred million years (Watson et al. 1982), but may have been throttled by oxygen accumulation (Wordsworth & Pierrehumbert 2014). Notably, massive water loss during a runaway greenhouse has been suggested as producing substantial $O_2$ in exoplanet atmospheres (Luger & Barnes 2015), but Venus serves as a counter-example to this. Hydration of surface rocks (Matsui & Abe 1986) or top-of-atmosphere loss processes (Chassefière 1997; Collinson et al. 2016) are potential sinks for water. Thus, Venus is an ideal laboratory to test hypotheses of abiotic oxygen loss processes.

Cloud-top variations of $SO_2$ have been observed across several decades from *Pioneer Venus* to *Venus Express* observations (Marcq et al. 2012), implying a long-term atmospheric cycling mechanism, or possibly to injections via volcanism. Recently, nine emissivity anomalies due to compositional differences were identified by the VEx Visible and Infrared Thermal Imaging Spectrometer (VIRTIS) aboard *Venux Express* as sites of potentially recent volcanism (Smrekar et al. 2010, S2010). There are purported lava flows associated with these anomalies estimated to be 2.5 million years old at most, and more likely to be as young as 250,000 years old or less (S2010) based on expected weathering rates of freshly emplaced basalts. The emissivity anomalies sit atop regions of thin, elastic lithosphere according to *Magellan* gravity data, strengthening the volcanism interpretation. In 2015, additional evidence for active volcanism on Venus was uncovered with a new analysis of *Venus Express'* Venus Monitoring Camera (VMC) data. Four temporally variable surface hotspots were discovered at the Ganiki Chasma rift zone near volcanoes Ozza Mons and Maat Mons (Shalygin et al., 2015), suggestive of present volcanic activity. However, interpreting these types of observations from above the cloud layer correctly is a challenge. The scattering footprint of radiation from the Venus surface escaping through the cloud deck is about 100 $km^2$, so smaller areas of increased thermal emission are smeared out.

## 3. Critical questions: The Need to Understand Earth's Twin

Many significant questions remain on the current state of Venus, suggesting major gaps in our understanding of the evolution of silicate planets, including the future evolution of Earth. Major outstanding questions include:
- Did Venus have a habitable period (e.g. Way et al. 2016)? That is, did Venus ever cool from a syn-accretionary runaway greenhouse?
- Where did the water go? Was hydrogen loss and abiotic oxygen production rampant, or did surface hydration dominate?
- What has the history of tectonics, volatile cycling, and volcanic resurfacing been? When did Venus enter its present stagnant-lid regime? Does any subduction occur today?

Venus accounts for 40% of the mass of terrestrial planets in our Solar system, yet even fundamental parameters such as the relative size of its core to mantle are unknown. As we expand the scope of planetary science to include those planets around other stars, the lack of measurements for basic planetary properties such as moment of inertia, core-size and state, seismic velocity and density variations with depth, and thermal profile for Venus hinders our ability to compare the potential uniqueness of the Earth and our Solar System to other planetary systems. Furthermore, the relative abundances of Venus' refractory elements can greatly inform the degree of mixing of planetesimals within this critical zone in the disk: where terrestrial planet are formed. If these relative refractory ratios are reflected in the size of its core, we gain by constraining even this simple structural parameter of Venus, a key benchmark in future studies of how our Solar system formed. This, in turn, will greatly aid in our studies of exoplanets, where stellar composition may set the initial compositional gradient of planetesimals within the disk but degree of mixing remains an elusive, unconstrained parameter.

## 4. A Plethora of Venus Analogs

The inner and outer boundaries of the HZ for various main sequence stars have been estimated using climate models, such as those by Kasting et al. (1993), and more recently by Kopparapu et al. (2013, 2014). An important aspect of these HZ calculations is that they provide a means to estimate the fraction of stars with Earth-size planets in the HZ, or eta-Earth. Much of the recent calculations of eta-Earth utilize *Kepler* results since these provide a large sample of terrestrial size objects from which to perform meaningful statistical analyses (Dressing & Charbonneau 2013, 2015; Kopparapu 2013; Petigura et al. 2013).

The transit method has a dramatic bias towards the detection of planets which are closer to the host star than farther away (Kane & von Braun 2008). Additionally, a shorter orbital period will result in an increased signal-to-noise (S/N) of the transit signature due to the increased number of transits observed within a given timeframe. The consequence of this is that *Kepler* has preferentially detected planets interior to the HZ which are therefore more likely to be potential Venus analogs than Earth analogs. Since the divergence of the Earth/Venusian atmospheric evolutions is a critical component for understanding Earth's habitability, the frequency of Venus analogs (eta-Venus) is also important to quantify.

Kane et al. (2014) defined the "Venus Zone" (VZ) as a target selection tool to identify terrestrial planets where the atmosphere could potentially be pushed into a runaway greenhouse producing surface conditions similar to those found on Venus. The below figure shows the VZ (red) and HZ (blue) for stars of different temperatures. The outer boundary of the VZ is the "Runaway Greenhouse" line which is calculated using climate models of Earth's atmosphere. The inner boundary (red dashed line) is estimated based on where the stellar radiation from the star would cause complete atmospheric erosion. The pictures of Venus shown in this region represent planet candidates detected by *Kepler*. Kane et al. (2014) calculated an occurrence rate of VZ terrestrial planets as 32% for low-mass stars and 45% for Sun-like stars. Note however that, like the HZ, the boundaries of the VZ should be considered a testable hypothesis since runaway greenhouse could occur beyond the calculated boundary (H2013, Foley 2015).

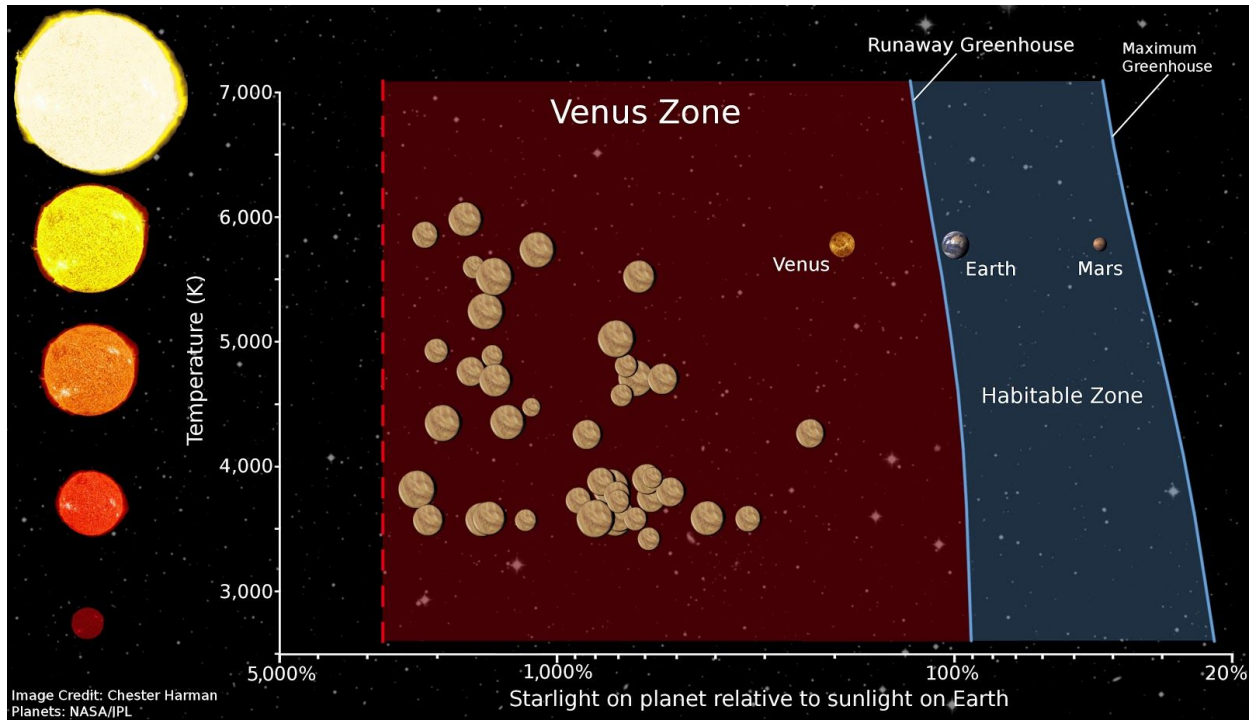

The prevalence of Venus analogs will continue to be relevant in the era of the *Transiting Exoplanet Survey Satellite (TESS)* mission, as hundreds of terrestrial planets orbiting bright host stars are expected to be detected (Sullivan et al. 2015). These will provide key opportunities for transmission spectroscopy follow-up observations using *JWST*, amongst other facilities. Such observations capable of identifying key atmospheric abundances for terrestrial planets will face the challenge of distinguishing between possible Venus and Earth-like surface conditions. Discerning the actual occurrence of Venus analogs will help us to decode why the atmosphere of Venus so radically diverged from its sister planet, Earth.

## 5. The Path Forward

The only in-situ terrestrial planet data available to us are here in our solar system. Thus, it is imperative that we gather improved information on Venus to aid in modeling both habitability and planetary interiors. The greatest advances in studies of Venus will come from a better understanding of the top-level questions described in Section 3 for which a series of missions - at multiple cost scales - could address parts thereof. Atmospheric modeling of exoplanets is of critical importance and an improved sampling of pressure, temperature, composition, and dynamics of the Venusian atmosphere as a function latitude would aid enormously in our ability to study exoplanetary atmospheres. In particular, new direct measurements of D/H within and below the clouds are needed to better constrain the volume of water present in Venus' history. Combined with D/H, isotopic measurements in the atmosphere would yield insights into the origins and fate of the Venusian atmosphere. A descent probe or lander to the surface (as a Discovery- or New Frontiers-class mission, or as part of a larger flagship mission) would make significant new measurements of atmospheric structure and D/H, as well as noble gas abundances and isotopic ratios. Such a mission could also provide first-ever measurements of the deepest atmosphere. Aerial platforms, such as balloons complement the vertical probe profiles by providing 2-D coverage of cloud region. These fundamental measurements would stimulate progress on multiple fronts, and vastly improve our understanding of both modern Venus and the transition of Venus to its modern state. Inclusion of a seismometer on future landers or long-term orbiters to measure moment of inertia, will provide new knowledge about the Venusian interior that is a critical, and necessary, step to expand our inferred knowledge of *any* exoplanet system.